\documentclass[intlimits,twoside,a4paper]{article}

\usepackage{wrapfig}\usepackage[T2A]{fontenc}
\usepackage[cp1251]{inputenc}

\usepackage{cmpj3}

\issue{2018}{21}{1}{13705}
\doinumber{10.5488/CMP.21.13705}

\title[Bethe lattice approach study of the mixed spin-$1/2$ and spin-$7/2$ Ising model]{ Bethe lattice approach study of the mixed spin-$\frac{1}{2}$ and spin-$\frac{7}{2}$ Ising model in a longitudinal 
magnetic field}

\author[S. Eddahri \textit{et al.}]{ S. Eddahri\refaddr{label4}, M. Karimou\refaddr{label1}, A. Razouk\refaddr{label4,label5}, F.~Hontinfinde\refaddr{label1,label2}\footnote{fhontinfinde@yahoo.fr}\,, 
 A. Benyoussef\refaddr{label3,label6} }
\addresses{
\addr{label4} Material Physics Laboratory, Faculty of Sciences and Technology, Sultan Moulay Slimane University, Morocco\\
\addr{label1} Institute of Mathematics and Physical Sciences (IMSP), Republic of Benin\\
\addr{label5} Department of Physics, Polydisciplinary Faculty, Sultan
Moulay Slimane University, Morocco\\ 
\addr{label2} University of Abomey-Calavi, Department of Physics, Republic of Benin\\
\addr{label3} LMPHE, Faculty of Sciences, Mohammed V University, Rabat, Morocco\\
\addr{label6} Hassan II Academy of Sciences and Technology, Rabat, Morocco}

\date{Received December 19, 2017, in final form February 7, 2018}

\begin{document}
\maketitle
\begin{abstract}

The magnetic properties of the mixed spin-$\frac{1}{2}$ and 
spin-$\frac{7}{2}$ Ising model with a crystal-field in a longitudinal magnetic field are investigated on  the Bethe 
lattice using exact  recursion relations. The ground-state phase diagram is constructed. The temperature-dependent one
is displayed in the case of uniform crystal-field on the  $(k_{\text{B}}T/|J|, D/|J|)$ plane in the absence of the external
constraint for lattice coordination numbers  $z = 3, 4$, $6$. The order parameters and 
corresponding response functions as well as the internal energy are calculated and examined in detail in order to 
feature the real nature of phase boundaries and corresponding temperatures. The thermal variations of the 
average magnetization are classified according to the N\'{e}el nomenclature.

\pacs 75.10.Hk, 75.10.Dg, 05.70.Fh, 05.50.+q

\end{abstract}
 
\section{Introduction}
The investigation of mixed Ising systems has been of great interest in statistical mechanics during the past decades \cite{la1,la2,la3,la4}.
This is due to the revelation of novel critical magnetic properties not detected during studies of their single-spin
counterparts. These systems are used to model ferrimagnetic materials  whose  properties are often needed in modern
sophisticated technologies, such as magnetic recording, storage and reading devices \cite{la5,la6,la7,la8,la9,la10,la11}.

 Theoretically, such systems have been studied by several statistical mechanical methods:
 re\-nor\-ma\-li\-za\-tion-group technique \cite{la12,la13}, mean-field 
approximation \cite{la14,la15,la16,la17,la18}, effective-field theory \cite{la19,la20,la21,la22,la23,la24},
 Monte Carlo simulations \cite{la25,la26,la27,la28,la29}. Recently, Jiang and Bai \cite{la30} have
studied the influence of an external longitudinal magnetic field on the magnetic properties of a mixed spin-1/2 and spin-3/2 Ising ferromagnetic/ferrimagnetic bilayer system. By means
of the effective-field theory, Essaoudi et al. \cite{la31} also studied  the same model using a  probability distribution
technique. This investigation revealed a remarkable influence of the field strength on the magnetic properties of this system.
Few exactly solved mixed-Ising models exist in the literature. For recent review on the subject, the reader
 should refer to references \cite{la32,la33,la34,la35,la36,la37,la38,la39,la40,la41}. Experimentally,
 the investigation of such systems has been performed for many years and has shown strong effects
of the external constraint on the physical properties of the system \cite{la42,la43}.

 In this paper, the Bethe lattice (BL) approach is used to examine the magnetic properties of the
mixed  spin-$\frac{1}{2}$ and spin-$\frac{7}{2}$ Ising model 
with equal crystal-field  in the presence of a longitudinal magnetic field. The exact recursion relations are calculated
considering contribution to the total partition function of the system from sites deep inside the lattice \cite{la44}. This 
work aims at the study of the effects on the phase boundaries of the competition between the two parameters of the system: the magnetic field and the crystal-field strengths.

The remainder of this paper is arranged as follows. In section \ref{secII}, the formulation of the model on the
BL is clarified. Also, the order-parameters, the corresponding response functions, the internal 
energy and free energy are expressed in terms of
recursion relations. In section \ref{secIII}, we  discuss in detail the numerical results.
Finally, in the last section we conclude.

\section{Formulation of the model on the BL\label{secII}}
\begin{figure}[!b]	
\begin{center}
\includegraphics[angle=0,width=0.55\textwidth]{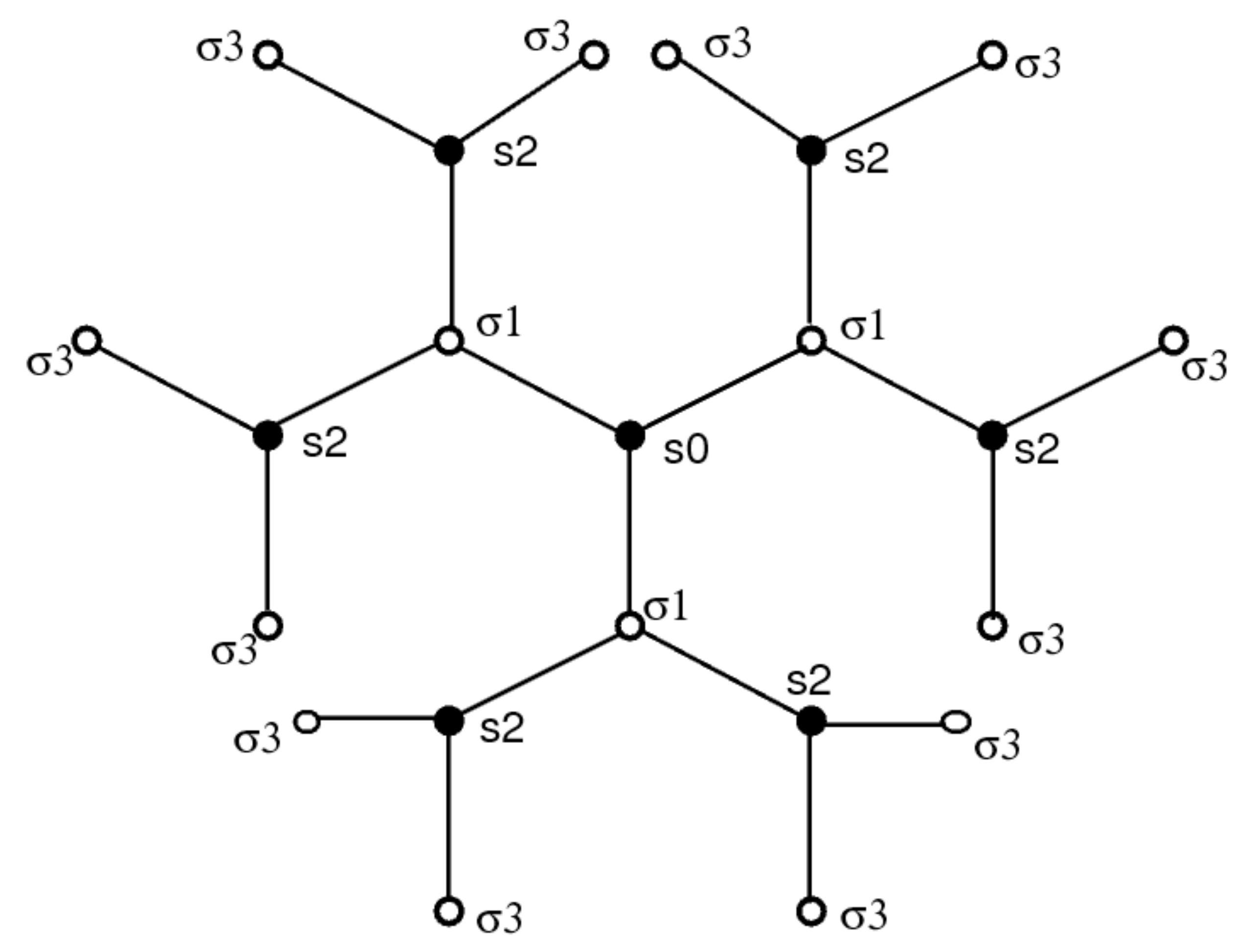}
\end{center}
\caption{Schematic representation of the Bethe lattice with coordination number $z=3$. It consists of
 two interpenetrating sublattices A and B with spin
variables $ s_{i} = 1/2$ and $\sigma_{j} = 7/2$, respectively.}
\label{figI}
\end{figure}

The mixed-spin system on the BL is shown in figure \ref{figI}. 
It consists of two sublattices $\text{A}$ and 
$\text{B}$. Sites of the sublattice $\text{A}$ are occupied by atoms of spins  $s_i=\pm{\frac{1}{2}}$. Those of 
the sublattice $\text{B}$ are occupied by atoms of spins  
$\sigma_j=\pm{\frac{7}{2}},\pm{\frac{5}{2}}, \pm{\frac{3}{2}},\pm{\frac{1}{2}}$. The BL
 is arranged such that the central spin is spin-$\frac{1}{2}$, and the next generation spin is 
spin-$\frac{7}{2}$ and so on to infinity. The Ising  Hamiltonian of the model  may be written as:
  \begin{eqnarray}
 H =-J \sum_{\langle{i,j}\rangle}{s_{i}}{\sigma_{j}} -D \sum_{j}{\sigma_{j}^{2}}-h\Big( \sum_{i}{s_{i}+ 
 \sum_{j}\sigma_{j}}\Big),
\end{eqnarray}
 where $J<0$ is the bilinear exchange coupling interaction strength; $D$ and $h$ are, respectively, 
 the crystal-field and the longitudinal magnetic field acting on the
spins.

In order to formulate the problem on the BL, the partition function is calculated.
Its expression reads:
\begin{eqnarray} 
Z= \sum{\exp{\bigg\{\beta \bigg[ J \sum_{<i,j>}{s_{i}}{\sigma_{j}} +D  \sum_{j}{\sigma_{j}^{2}} +
h\Big(\sum_{i}{s_{i}} + \sum_{j}{\sigma{_j}}\Big) \bigg]\bigg\}}}.
\end{eqnarray}

If the BL is cut at the central spin $s_0$, it splits into $z$  disconnected pieces. Then, 
the partition function can be written as:
\begin{eqnarray}
 Z=\sum_{s_0}{\exp{\big[\beta (h s_{0})\big]} g_{n}^{z}(s_0)}\,,
\end{eqnarray}
where $s_{0}$ is the central spin value of the lattice, $ g_{n}(s_{0})$ is the partition function of an individual branch and the suffix $n$
represents the fact that the sub-tree has $n$ shells, i.e., $n$ steps from the root to  the boundary sites.
If one continues to cut the BL on spins $\sigma_1$ and $s_2$ which are respectively
the nearest and next-nearest neighbors of the central spin $s_0$, the recurrence relations
for $ g_{n}(s_{0})$ and $g_{n-1}(\sigma_{1})$ read:
\begin{equation}
 g_{n}(s_{0})= \sum_{\{\sigma_{1}\}}\exp\big[\beta(Js_{0}\sigma_{1}+D\sigma_{1}^{2}+
 h \sigma_1)\big]\big[g_{n-1}(\sigma_{1})\big]^{z-1},
\end{equation}
\begin{equation}
 g_{n-1}(\sigma_{1})= \sum_{\{s_{2}\}}\exp\big[\beta(Js_{2}\sigma_{1}+h s_{2})\big]\big[g_{n-2}(s_{2})\big]^{z-1}.
\end{equation}

Explicit relations for some  $g_{n}(s_{0})$ and $g_{n-1}(\sigma_{1})$ are given in
 the following:
 \begin{eqnarray}
    g_{n}\left(\pm\frac{1}{2}\right)&=& \sum_{\{\sigma_{1}\}}\exp\bigg[\beta\Big(\pm \frac{J}{2} \sigma_{1}+D\sigma_{1}^{2}+
    h \sigma_1\Big)\bigg]\big[g_{n-1}(\sigma_{1})\big]^{z-1}\nonumber\\
   &=& \exp{\bigg[\beta\Big(\pm\frac{7J}{4}+\frac{49}{4}D +\frac{7}{2}h\Big)\bigg]}\bigg[g_{n-1}\left(\frac{7}{2}\right)\bigg]^{z-1}\nonumber\\
  &+& \exp{\bigg[\beta\Big(\mp\frac{7J}{4} +\frac{49}{4}D -\frac{7}{2}h \Big)\bigg]}\bigg[g_{n-1}\left(-\frac{7}{2}\right)\bigg]^{z-1}\nonumber\\
   &+& \exp{\bigg[\beta\Big(\pm\frac{5J}{4}+\frac{25}{4}D +\frac{5}{2}h\Big)\bigg]}\bigg[g_{n-1}\left(\frac{5}{2}\right)\bigg]^{z-1}\nonumber\\
  &+& \exp{\bigg[\beta\Big(\mp\frac{5J}{4} +\frac{25}{4}D -\frac{5}{2}h \Big)\bigg]}\bigg[g_{n-1}\left(-\frac{5}{2}\right)\bigg]^{z-1}\nonumber\\
   &+& \exp{\bigg[\beta\Big(\pm\frac{3J}{4} +\frac{9}{4}D +\frac{3}{2}h\Big)\bigg]}\bigg[g_{n-1}\left(\frac{3}{2}\right)\bigg]^{z-1}\nonumber\\
   &+&  \exp{\bigg[\beta\Big(\mp\frac{3J}{4} +\frac{9}{4}D -\frac{3}{2}h\Big)\bigg]}\bigg[g_{n-1}\left(-\frac{3}{2}\right)\bigg]^{z-1} \nonumber \\
  &+&  \exp{\bigg[\beta\Big(\pm\frac{J}{4}+\frac{1}{4}D +\frac{1}{2}h\Big)\bigg]}\bigg[g_{n-1}\left(\frac{1}{2}\right)\bigg]^{z-1}\nonumber\\
  &+& \exp{\bigg[\beta\Big(\mp\frac{J}{4} +\frac{1}{4}D -\frac{1}{2}h\Big)\bigg]}\bigg[g_{n-1}\left(-\frac{1}{2}\right)\bigg]^{z-1},  
  \end{eqnarray}
  
   \begin{eqnarray}
    g_{n-1}\left(\pm\frac{7}{2}\right)&=& \sum_{\{s_{2}\}}\exp\bigg[\beta\Big(\pm\frac{7J}{2}{s_{2}}+
    h s_{2}\Big)\bigg]\big[g_{n-2}(s_{2})\big]^{z-1}\nonumber\\
  &=& \exp{\bigg[\beta\Big(\pm\frac{7J}{4}+\frac{h}{2}\Big)\bigg]}\bigg[g_{n-2}\left(\frac{1}{2}\right)\bigg]^{z-1}\nonumber\\ 
 &+& \exp{\bigg[\beta\Big(\mp\frac{7J}{4}-\frac{h}{2}\Big)\bigg]}\bigg[g_{n-2}\left(-\frac{1}{2}\right)\bigg]^{z-1},  
 \end{eqnarray}
  \begin{eqnarray}
    g_{n-1}\left(\pm\frac{5}{2}\right)&=& \sum_{\{s_{2}\}}\exp\bigg[\beta\Big(\pm\frac{5J}{2}{s_{2}}+
    h s_{2}\Big)\bigg]\big[g_{n-2}(s_{2})\big]^{z-1}\nonumber\\
  &=& \exp{\bigg[\beta\Big(\pm\frac{5J}{4}+\frac{h}{2}\Big)\bigg]}\bigg[g_{n-2}\left(\frac{1}{2}\right)\bigg]^{z-1}\nonumber\\ 
 &+& \exp{\bigg[\beta\Big(\mp\frac{5J}{4}-\frac{h}{2}\Big)\bigg]}\bigg[g_{n-2}\left(-\frac{1}{2}\right)\bigg]^{z-1},  
 \end{eqnarray} 
  \begin{eqnarray}
    g_{n-1}\left(\pm\frac{3}{2}\right)&=& \sum_{\{s_{2}\}}\exp\bigg[\beta\Big(\pm\frac{3J}{2}{s_{2}}+
    h s_{2}\Big)\bigg]\big[g_{n-2}(s_{2})\big]^{z-1}\nonumber\\
  &=& \exp{\bigg[\beta\Big(\pm\frac{3J}{4}+\frac{h}{2}\Big)\bigg]}\bigg[g_{n-2}\left(\frac{1}{2}\right)\bigg]^{z-1}\nonumber\\ 
 &+& \exp{\bigg[\beta\Big(\mp\frac{3J}{4}-\frac{h}{2}\Big)\bigg]}\bigg[g_{n-2}\left(-\frac{1}{2}\right)\bigg]^{z-1}, 
 \end{eqnarray}
  \begin{eqnarray}
    g_{n-1}\left(\pm\frac{1}{2}\right)&=& \sum_{\{s_{2}\}}\exp\bigg[\beta\Big(\pm\frac{1J}{2}{s_{2}}+
    h s_{2}\Big)\bigg]\big[g_{n-2}(s_{2})\big]^{z-1}\nonumber\\
  &=& \exp{\bigg[\beta\Big(\pm\frac{1J}{4}+\frac{h}{2}\Big)\bigg]}\bigg[g_{n-2}\left(\frac{1}{2}\right)\bigg]^{z-1}\nonumber\\ 
 &+& \exp{\bigg[\beta\Big(\mp\frac{1J}{4}-\frac{h}{2}\Big)\bigg]}\bigg[g_{n-2}\left(-\frac{1}{2}\right)\bigg]^{z-1}.  
 \end{eqnarray}
 After calculating all the $g_{n}(s_{0})$ and $g_{n-1}(\sigma_{1})$, the recursion 
 relations for the spin-$\frac{1}{2}$ are defined as:
\begin{equation}
  Y_{n}= \frac{g_{n}\big(\frac{1}{2}\big)}{g_{n}\big(-\frac{1}{2}\big)} 
\end{equation}
and for the spin-$\frac{7}{2}$ as:
\begin{eqnarray}
A_{n-1}= \frac{g_{n-1}\big(+\frac{7}{2}\big)}{g_{n-1}\big(-\frac{1}{2}\big)}\,, \qquad B_{n-1}= \frac{g_{n-1}\big(-\frac{7}{2}\big)}{g_{n-1}\big(-\frac{1}{2}\big)}\,,\\
 C_{n-1}= \frac{g_{n-1}\big(+\frac{5}{2}\big)}{g_{n-1}\big(-\frac{1}{2}\big)}\,, \qquad D_{n-1}= \frac{g_{n-1}\big(-\frac{5}{2}\big)}{g_{n-1}\big(-\frac{1}{2}\big)}\,,\\ 
  E_{n-1}= \frac{g_{n-1}\big(\frac{3}{2}\big)}{g_{n-1}\big(-\frac{1}{2}\big)}\,, \qquad  F_{n-1}= \frac{g_{n-1}\big(-\frac{3}{2}\big)}{g_{n-1}\big(-\frac{1}{2}\big)}\,,\nonumber\\
   G_{n-1}= \frac{g_{n-1}\big(+\frac{1}{2}\big)}{g_{n-1}\big(-\frac{1}{2}\big)}\,. \nonumber
\end{eqnarray}
For the numerical investigation of the model, the magnetization $M$ and the corresponding quadrupolar moment $Q$ are
 quantities of interest. For the 
sublattice A, the sublattice magnetization $M_{1/2}$ is defined by:
\begin{eqnarray}
 M_{1/2} = Z_{1/2}^{-1} \sum_{\{s_{0}\}}s_{0}{\exp{(\beta h s_{0})} g_{n}^{z}(s_0)}.
 \end{eqnarray}
After some mathematical manipulations, the sublattice magnetization $M_{1/2}$ is explicitly given by:
\begin{eqnarray}
 M_{1/2} =  \frac{\exp{\big(\frac{\beta h}{2}\big)}Y_{n}^{z}-\exp{\big(-\frac{\beta h}{2}\big)}}{2\big[\exp{\big(\frac{\beta h}{2}\big)}Y_{n}^{z}+\exp{\big(-\frac{\beta h}{2}\big)}\big]}.
\end{eqnarray}
In the same way, the two order-parameters for the sublattice B are calculated as 
follows:
\begin{eqnarray}
  M_{7/2}=\frac{M_{7/2}^{\prime}}{M_{7/2}^{0}}\,, \qquad 
 Q_{7/2}=\frac{Q_{7/2}^{\prime}}{Q_{7/2}^{0}},
 \end{eqnarray} 
where
\begin{eqnarray}
 M_{7/2}^{\prime} &=& 7\exp{\left(\frac{49}{4}\beta D\right)}\bigg[\exp{\left(\frac{7}{2}\beta h\right)}A_{n-1}^{z}-
 \exp{\left(-\frac{7}{2}\beta h\right)}B_{n-1}^{z}\bigg]\nonumber\\
 &+& 5\exp{\left(\frac{25}{4}\beta D\right)}\bigg[\exp{\left(\frac{5}{2}\beta h\right)}C_{n-1}^{z}-
 \exp{\left(-\frac{5}{2}\beta h\right)}D_{n-1}^{z}\bigg]\nonumber\\
 &+& 3\exp{\left(\frac{9}{4}\beta D\right)}\bigg[\exp{\left(\frac{3}{2}\beta h\right)}E_{n-1}^{z}-
 \exp{\left(-\frac{3}{2}\beta h\right)}F_{n-1}^{z}\bigg] \nonumber\\
 &+&\exp{\left(\frac{1}{4}\beta D\right)}\bigg[\exp{\left(\frac{1}{2}\beta h\right)}G_{n-1}^{z}-
 \exp{\left(-\frac{1}{2}\beta h\right)}\bigg], 
\end{eqnarray}

\begin{eqnarray}
M_{7/2}^{0} &=& 2\exp{\left(\frac{49}{4}\beta D\right)}\bigg[\exp{\left(\frac{7}{2}\beta h\right)}A_{n-1}^{z}
+\exp{\left(-\frac{7}{2}\beta h\right)}B_{n-1}^{z}\bigg]\nonumber\\
&+& 2\exp{\left(\frac{25}{4}\beta D\right)}\bigg[\exp{\left(\frac{5}{2}\beta h\right)}C_{n-1}^{z}
+\exp{\left(-\frac{5}{2}\beta h\right)}D_{n-1}^{z}\bigg]\nonumber\\
 &+& 2\exp{\left(\frac{9}{4}\beta D\right)}\bigg[\exp{\left(\frac{3}{2}\beta h\right)}E_{n-1}^{z}+
 \exp{\left(-\frac{3}{2}\beta h\right)}F_{n-1}^{z}\bigg] \nonumber\\
&+& 2\exp{\left(\frac{1}{4}\beta D\right)}\bigg[\exp{\left(\frac{1}{2}\beta h\right)}G_{n-1}^{z}+
\exp{\left(-\frac{1}{2}\beta h\right)}\bigg], 
\end{eqnarray}

\begin{eqnarray}
Q_{7/2}^{\prime} &=& 49\exp{\left(\frac{49}{4}\beta D\right)}\bigg[\exp{\left(\frac{7}{2}\beta h\right)}A_{n-1}^{z} +
\exp{\left(-\frac{7}{2}\beta h\right)}B_{n-1}^{z}\bigg]\nonumber\\
 &+& 25\exp{\left(\frac{25}{4}\beta D\right)}\bigg[\exp{\left(\frac{5}{2}\beta h\right)}C_{n-1}^{z} +
\exp{\left(-\frac{5}{2}\beta h\right)}D_{n-1}^{z}\bigg]\nonumber\\
 &+& 9\exp{\left(\frac{9}{4}\beta D\right)}\bigg[\exp{\left(\frac{3}{2}\beta h\right)}E_{n-1}^{z}+\exp{\left(-\frac{3}{2}\beta h\right)}F_{n-1}^{z}\bigg] \nonumber\\
 &+&\exp{\left(\frac{1}{4}\beta D\right)}\bigg[\exp{\left(\frac{1}{2}\beta h\right)}G_{n-1}^{z}+\exp{\left(-\frac{1}{2}\beta h\right)}\bigg], 
\end{eqnarray}
\begin{eqnarray}
Q_{7/2}^{0} &=& 4\exp{\left(\frac{49}{4}\beta D\right)}\bigg[\exp{\left(\frac{7}{2}\beta h\right)}A_{n-1}^{z}+
\exp{\left(-\frac{7}{2}\beta h\right)}B_{n-1}^{z}\bigg]\nonumber\\
&=& 4\exp{\left(\frac{25}{4}\beta D\right)}\bigg[\exp{\left(\frac{5}{2}\beta h\right)}C_{n-1}^{z}+
\exp{\left(-\frac{5}{2}\beta h\right)}D_{n-1}^{z}\bigg]\nonumber\\
 &+& 4\exp{\left(\frac{9}{4}\beta D\right)}\bigg[\exp{\left(\frac{3}{2}\beta h\right)}E_{n-1}^{z}+\exp{\left(-\frac{3}{2}\beta h\right)}F_{n-1}^{z}\bigg] \nonumber\\
 &+& 4\exp{\left(\frac{1}{4}\beta D\right)}\bigg[\exp{\left(\frac{1}{2}\beta h\right)}G_{n-1}^{z}+\exp{\left(-\frac{1}{2}\beta h\right)}\bigg].
\end{eqnarray}

 In order to determine the compensation temperature, one should define the global magnetization $M_{\text{net}}$ of the model
which is given by:
\begin{eqnarray}
 M_{\text{net}}=\frac{M_{1/2}+M_{7/2}}{2}.
\end{eqnarray}
To really study the model in detail and  single out the influence of the crystal-field and the applied magnetic field 
on the magnetic properties of the model, we have also examined the thermal variations of the response
functions i.e., the susceptibilities, the specific heat and the internal energy defined respectively by:
\begin{eqnarray}
 \chi_\text{Total}=\chi_{1/2} + \chi_{7/2}
 =\left(\frac{\partial{M_{1/2}}}{\partial{h}} \right)_{h=0} + \left(\frac{\partial{M_{7/2}}}{\partial{h}} \right)_{h=0},
\end{eqnarray}
\vspace{-6mm}
\begin{eqnarray}
  C = -{\beta}^{2}\frac{{\partial}^{2}{(-\beta F')}}{\partial{{\beta}^{2}}}\,,
\end{eqnarray}
\begin{eqnarray}
 \frac{U}{N|J|}=-k_\text{B}T^{2}\frac{\partial}{\partial T}\left(\frac{F^{\prime}}{k_\text{B}T}\right),
\end{eqnarray}
where $ F^{\prime} $ is the free energy of the model.

In order to classify the order of the phase transitions, i.e., whether the second- or the first-order, the free energy
expression is also needed in addition to the order-parameters. It can be calculated in terms of recurrence
relations by using the definition  $ F^{\prime} = -kT\ln(Z)$ in the thermodynamic limit, i.e.,
 $n\rightarrow\infty$ : 
  \begin{eqnarray}
   \frac{F^{\prime}}{J} &=& -\frac{1}{{\beta}^{'}}\Bigg(\frac{z-1}{2-z}\ln\bigg\{\exp{\left[\beta\left(-\frac{J}{4} +\frac{h}{2}\right)\right]}Y_{n}^{z-1}
   +\exp{\left[\beta\left(\frac{J}{4}-\frac{h}{2}\right)\right]}\bigg\}\Bigg) \nonumber\\
   &-& \frac{1}{{\beta}^{'}}\Bigg(\ln\bigg\{\exp{\left[\beta\left(\frac{h}{2} \right)\right]}Y_{n}^{z} +\exp{\left[\beta\left(-\frac{h}{2}\right)\right]}\bigg\}\Bigg) \nonumber\\
   &-& \frac{1}{{\beta}^{'}}\Bigg(\frac{1}{2-z}\ln\bigg\{\exp{\left[\beta\left(-\frac{7J}{4} +
   \frac{49 D}{4} +\frac{7h}{2}\right)\right]}A_{n-1}^{z-1} \nonumber\\
   &+& \exp{\left[\beta\left(\frac{7J}{4} +\frac{49D}{4} -\frac{7h}{2}\right)\right]}B_{n-1}^{z-1} \nonumber \\
   &+& \exp{\left[\beta\left(-\frac{5J}{4} +\frac{25 D}{4} +\frac{5h}{2}\right)\right]}C_{n-1}^{z-1} +
   \exp{\left[\beta\left(\frac{5J}{4} +\frac{25D}{4} -\frac{5h}{2}\right)\right]}D_{n-1}^{z-1} \nonumber \\
   &+& \exp{\left[\beta\left(-\frac{3J}{4} +\frac{9D}{4} +\frac{3h}{2} \right)\right]}E_{n-1}^{z-1} +
   \exp{\left[\beta\left(\frac{3J}{4} +\frac{9D}{4} -\frac{3h}{2} \right)\right]}F_{n-1}^{z-1} \nonumber \\
   &+& \exp{\left[\beta\left(-\frac{J}{4} +\frac{D}{4} +\frac{h}{2}\right)\right]}G_{n-1}^{z-1} +
   \exp{\left[\beta\left(\frac{J}{4} +\frac{D}{4} -\frac{h}{2}\right)\right]}\bigg\}\Bigg).
 \end{eqnarray}
 
  \section{Numerical results and discussions\label{secIII}}
  
  In this section, we present and discuss the results we obtained for the temperature
  phase diagrams of the model, the thermal variations of the order-parameters, the response functions 
  and the internal energy. To this end, we first construct  the  phase diagram at $T=0$.
  \subsection{Ground-state phase diagrams}
  
\begin{figure}[!t]
\begin{center}
	\vspace{-4mm}
\includegraphics[angle=0,width=0.69\textwidth]{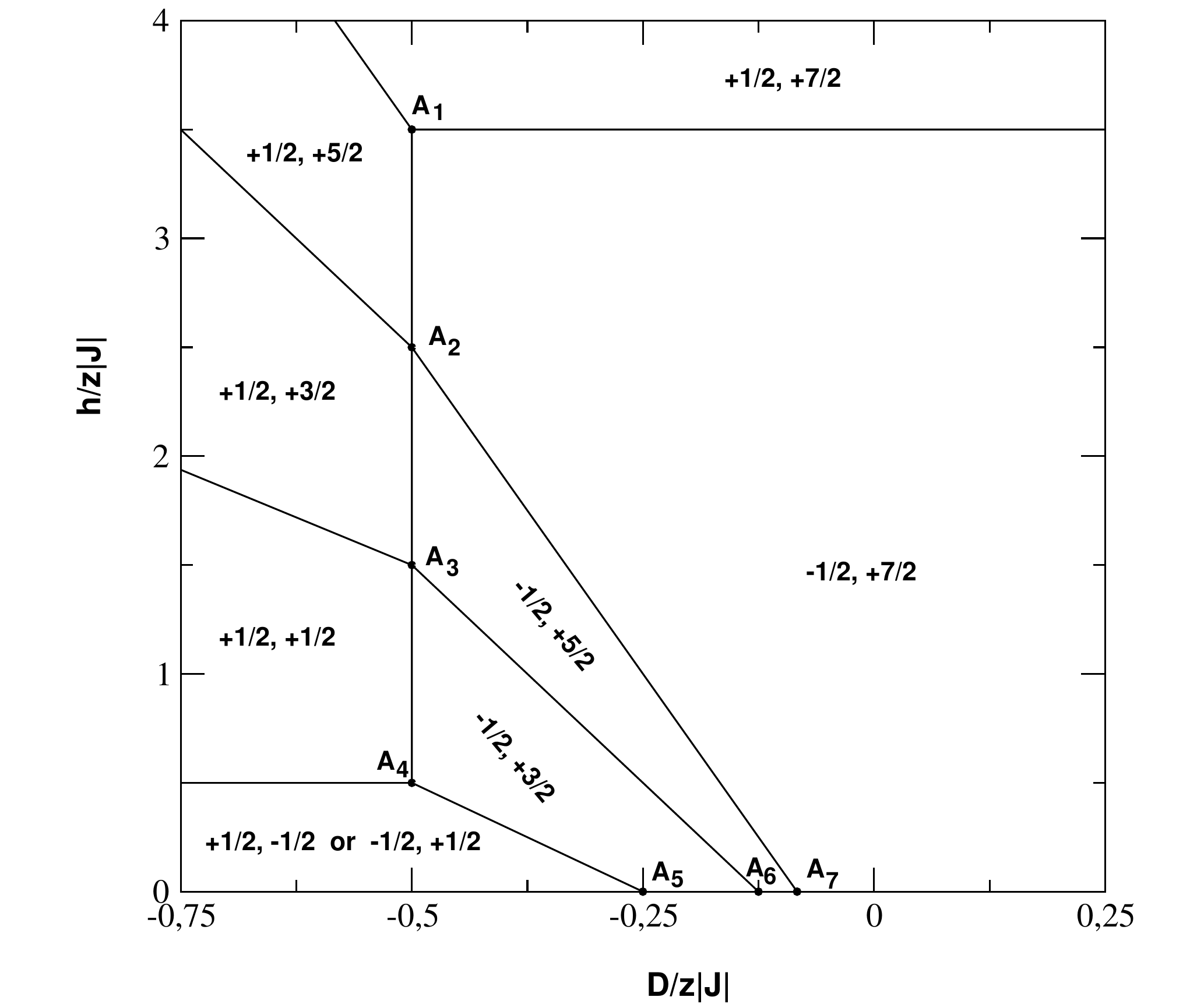}
\end{center}
\vspace{-2mm}
\caption{ Ground-state phase diagram of the mixed spin-$\frac{1}{2}$ and spin-$\frac{7}{2}$ Ising ferrimagnetic model with 
the crystal-field $D$ applied  one sublattice  in the $\left(h/z|J |, D/z|J |\right)$ plane. There exist eight stable phases.
Along the $D/z|J |$-axis and for all values of $z$, three hybrid phases may appear at the multicritical
points $A_{5}$, $A_{6}$
and $A_{7}$.}
\label{figure2}
\end{figure}

   The ground-state phase diagram of the model is obtained by comparing the values of the energy $E_0$  
    for different spin configurations which can be expressed as:
  \begin{eqnarray}
   E_0=s\sigma-\frac{1}{z |J|}\left[D  {\sigma}^{2}+h (s+\sigma)\right].
  \end{eqnarray}
 Only  eight possible pairs of spins due to the ferrimagnetic coupling $J$ and positive field $(h \geqslant 0)$ are
 found.
Calculations of these energies in the $(h/z|J|, D/z|J|)$ plane yield the ground-state
phase diagram displayed in figure~\ref{figure2}. The  model
has a usual spin-flip symmetry. Thus, all ground-states for negative field ($h<0$) can be obtained from the corresponding
 ones at a positive field, simply by reversing all spin orientations. Some key
features of the model are revealed in the diagram, in particular, the existence of seven multicritical points $(A_1, A_2,\cdots , A_{7})$ and
coexistence lines where spin pair energies of some phases are equal. In the absence of the magnetic field, for 
a given values of $z$ and $D/z|J|$,  $M_{7/2}$ shows seven saturation values whereas for $M_{1/2}$, $\pm{\frac{1}{2}}$  
are the two saturation values. Hence, we get the ferrimagnetic phases: $\mathbf{F} \big(\mp{\frac{1}{2}}, \pm{\frac{7}{2}}\big)$,
 $\mathbf{F} \big(\mp{\frac{1}{2}}, \pm{\frac{5}{2}}\big)$, $\mathbf{F} \big(\mp{\frac{1}{2}}, \pm{\frac{3}{2}}\big) $, $\mathbf{F} \big(\mp{\frac{1}{2}},
 \pm{\frac{1}{2}}\big) $ and at the borders of these phases, three hybrid phases: $\mathbf{F} \big(\mp{\frac{1}{2}}, \pm{1}\big)$,
$\mathbf{F} \big(\mp{\frac{1}{2}}, \pm{2}\big)$, $\mathbf{F} \big(\mp{\frac{1}{2}}, \pm{3}\big)$ at the multicritical points $A_5$, $A_6$ 
and $A_7$, respectively. These hybrid phases should correspond to cases
where the sublattice B is half-half covered by spins of the two neighboring phases.

 \subsection{Finite-temperature phase diagrams}
 
\begin{figure}[!t]
  	\vspace{-3mm}
 	\begin{center}
 		\includegraphics[angle=0,width=0.72\textwidth]{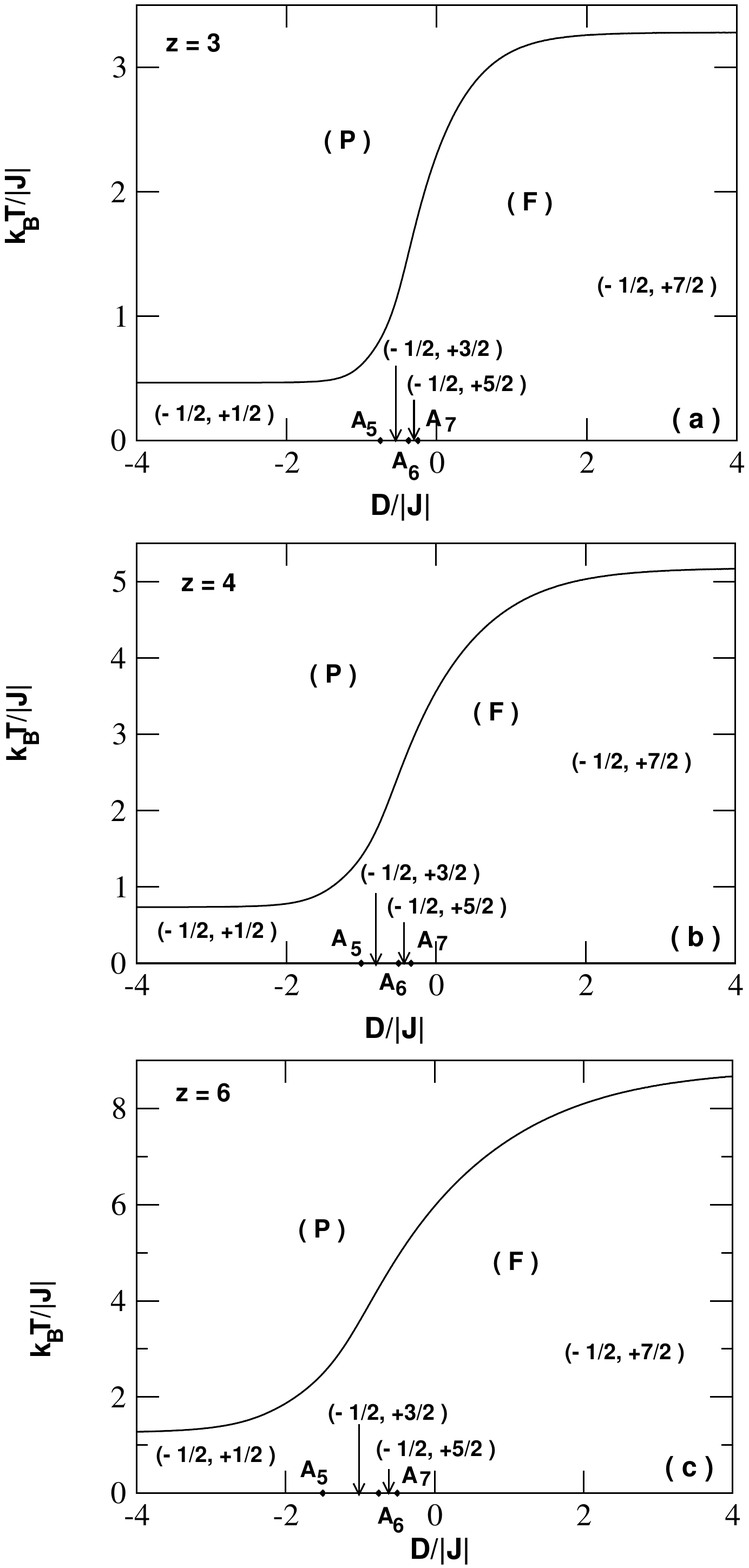}
 	\end{center}
 \vspace{-3mm}
 	\caption{Phase diagrams of the model in the $(D/|J |, k_\text{B}T/|J |)$ plane. The black circles on the $D/|J|$-axis 
 		correspond to the
 		$A_{5}$, $A_{6}$ and $A_{7}$ multicritical points obtained in the ground-state  phase diagram and the solid line indicates
 		the second-order transition line. Panel (a): $z = 3$; panel (b): $z = 4$ and  panel (c): $z = 6$.
 		Here, the model only presents second-order transition for all values of $z$. The multicritical points $A_{5}$, $A_{6}$ and
 		$A_{7}$ which respectively indicate the positions of the hybrid phases $F \big( \mp \frac{1}{2} , \pm 3 \big)$, $F \big( \mp \frac{1}{2} , \pm 2 \big)$,
 		and $F \big( \mp \frac{1}{2} , \pm 1 \big)$  respectively,
 		separate the ferrimagnetic phases $F \big(\mp \frac{1}{2} , \pm \frac{7}{2}\big)$, $F \big( \mp \frac{1}{2} ,\pm \frac{5}{2}\big)$, $F \big( \mp \frac{1}{2} , \pm \frac{3}{2})$,  and 
 		$F \big( \mp \frac{1}{2} , \pm \frac{1}{2}\big)$.}
 	\label{figure3}
 \end{figure}

 In this subsection, we  show some typical results for the mixed spin-$\frac{1}{2}$ and spin-$\frac{7}{2}$ Ising
 model on the BL with a crystal field at zero longitudinal magnetic field. First, we present phase diagrams of the model
  in the $(D/|J|, k_\text{B}T/|J|)$ plane for $z=3, 4$ and $6$. Therein,  
 solid lines indicate a second-order transition. The three filled circles $A_5$, $A_6$ and $A_7$ in figure~\ref{figure3} are the
 multicritical points found in the ground-state phase diagram.
  
 From this figure, some interesting properties of the system emerge. Indeed, for all values of 
  the coordination number $z$, from panel (a) to panel (c), transition
  lines are only of the second-order type and separate the ferrimagnetic phase $(\bf F)$ which is a mixture of
  five different ferrimagnetic phases from the paramagnetic phase $(P)$. They 
  become constant for $D/|J| <-\frac{z}{12}$. One  observes that for 
  $D/|J|>-\frac{z}{12}$, the second-order phase transition turns from ferrimagnetic phase 
  $\mathbf{F} \big(\mp \frac{1}{2}, \pm \frac{7}{2} \big)$ to the 
  disordered paramagnetic phase $P$. For $-\frac{z}{8}< D/|J|< -\frac{z}{12}$, the second-order
  phase transition is from the ferrimagnetic $\mathbf{F} \big(\mp \frac{1}{2}, \pm \frac{5}{2}\big)$  to the paramagnetic phase $P$.
   When $-\frac{z}{4}< D/|J|< -\frac{z}{8}$, the second-order phase transition is from the ferrimagnetic
  $\mathbf{F} \big(\mp \frac{1}{2}, \pm \frac{3}{2}\big)$  to the paramagnetic phase $P$. In the case where  
 $ D/|J|< -\frac{z}{4}$, the second-order phase transition is from the ferrimagnetic 
 phase $\mathbf{F} \big(\pm \frac{1}{2}, \mp \frac{1}{2}\big)$  to the paramagnetic phase $P$. For $D/|J|=-\frac{z}{12}$, 
  (respectively  $D/|J|=-\frac{z}{8}$ and $D/|J|=-\frac{z}{4}$ ), the 
  second-order transition phase is from the hybrid phase $\mathbf{F} \big(\mp{\frac{1}{2}}, \pm{3}\big)$ \big[respectively the hybrid
  phase $\mathbf{F} \big(\mp{\frac{1}{2}}, \pm{2}\big)$ and $\mathbf{F} \big(\mp{\frac{1}{2}}, \pm{1}\big)$\big]  to the paramagnetic phase $P$. 
  
  It is important to mention that figure~\ref{figure3} presents some resemblances with results from  
  references \cite{la24, la36, la40} concerning the second-order transition lines. Also, by increasing the value of the coordination number $z$,
  the ferrimagnetic domain $\bf F$ becomes important.

 \begin{figure}[!b]
	\begin{center}
		\includegraphics[angle=0,width=0.7\textwidth]{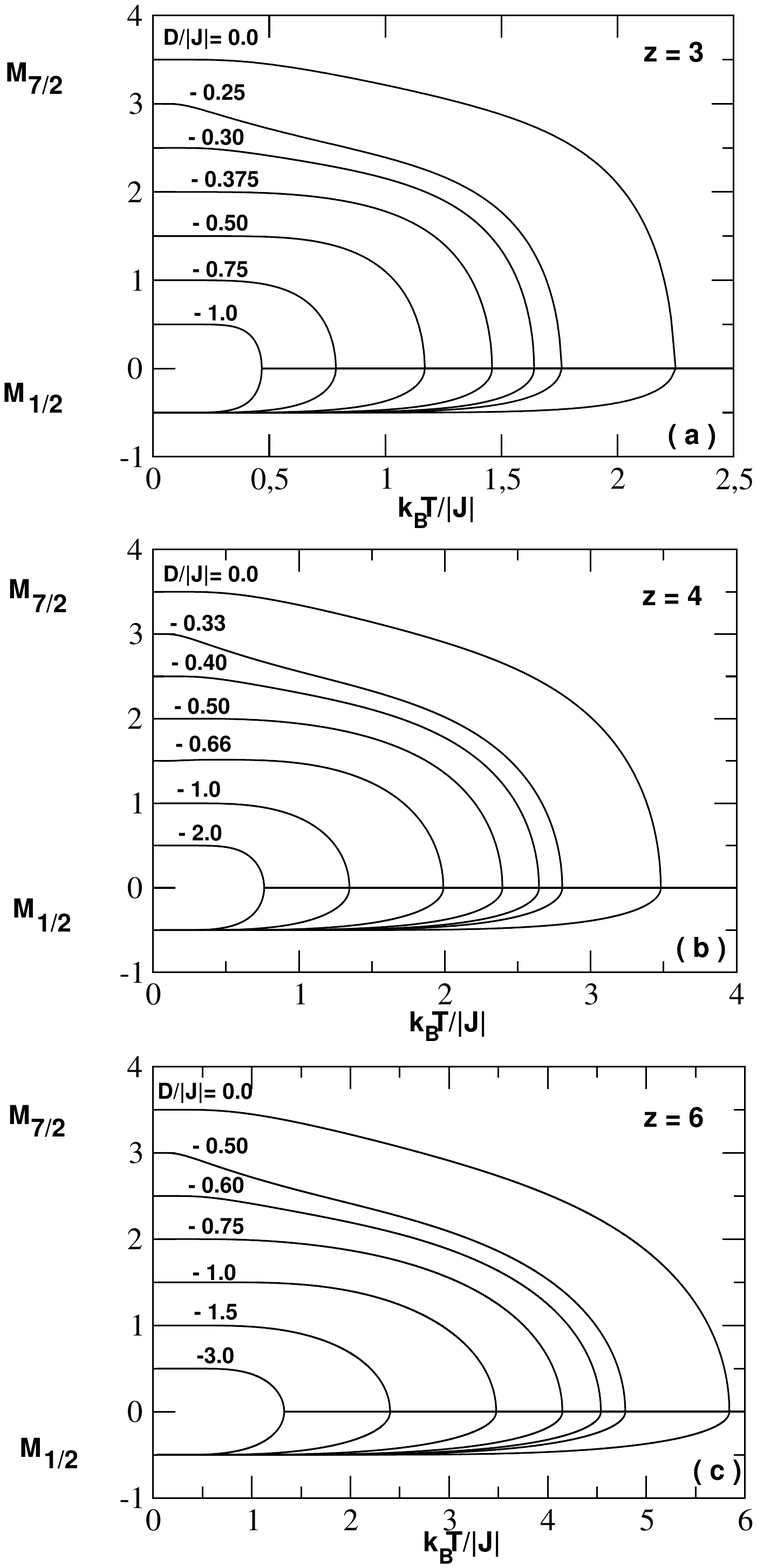}
	\end{center}
    \vspace{-2mm}
	\caption{ Sublattice magnetizations of the model as functions of the reduced temperature $k_{\text{B}}T/|J|$ for
		$z=3, 4$ and $6$ for 
		various values of the crystal-field interactions $D$. Panel (a): curves are displayed
		for $z=3$ and selected values of $D/|J|$ are indicated on the curves. Panel (b): curves are 
		displayed for $z=4$ and selected values of $D/|J|$ indicated on the curves. Panel (c): curves
		are displayed for $z=6$ and selected values of $D/|J|$ indicated on the curves. For all values of $z$, 
		the sublattice magnetization curves are all continuous.}
	\label{figure4}
\end{figure}

\subsection{Thermal variations of the order-parameters, the response functions and the internal energy}
 
 As it is explained above, the thermal variations of the order-parameters, the response
 functions and the internal energy
 for the  model were calculated in terms of recursion relations. The thermal 
 variations of the order-parameters are crucial for obtaining the temperature dependence phase
 diagrams of the model. In fact, when the magnetization curves go to zero continuously, one gets a second-order phase 
 transition.
 In the case of a jump in the magnetizations curves followed by a discontinuity of the first derivative
 of the free-energy $F'$, a first-order transition temperature is got.
 Besides these two temperatures, there is another temperature called 
 compensation temperature defined as the temperature where the global 
 magnetization becomes zero before the critical temperature. Therefore, in order to 
 identify transition and compensation lines, one needs to 
 study the thermal behaviours of the considered thermodynamical quantities of the model. Now, we  present 
 some results on 
 the thermal behaviours of the order-parameters, the response functions and the internal energy in the 
 the absence of the magnetic field $h$ when $z=3, 4$ and $6$. 
  \begin{figure}[!b]
  	\vspace{-3mm}
	\begin{center}
		\includegraphics[angle=0,width=0.695\textwidth]{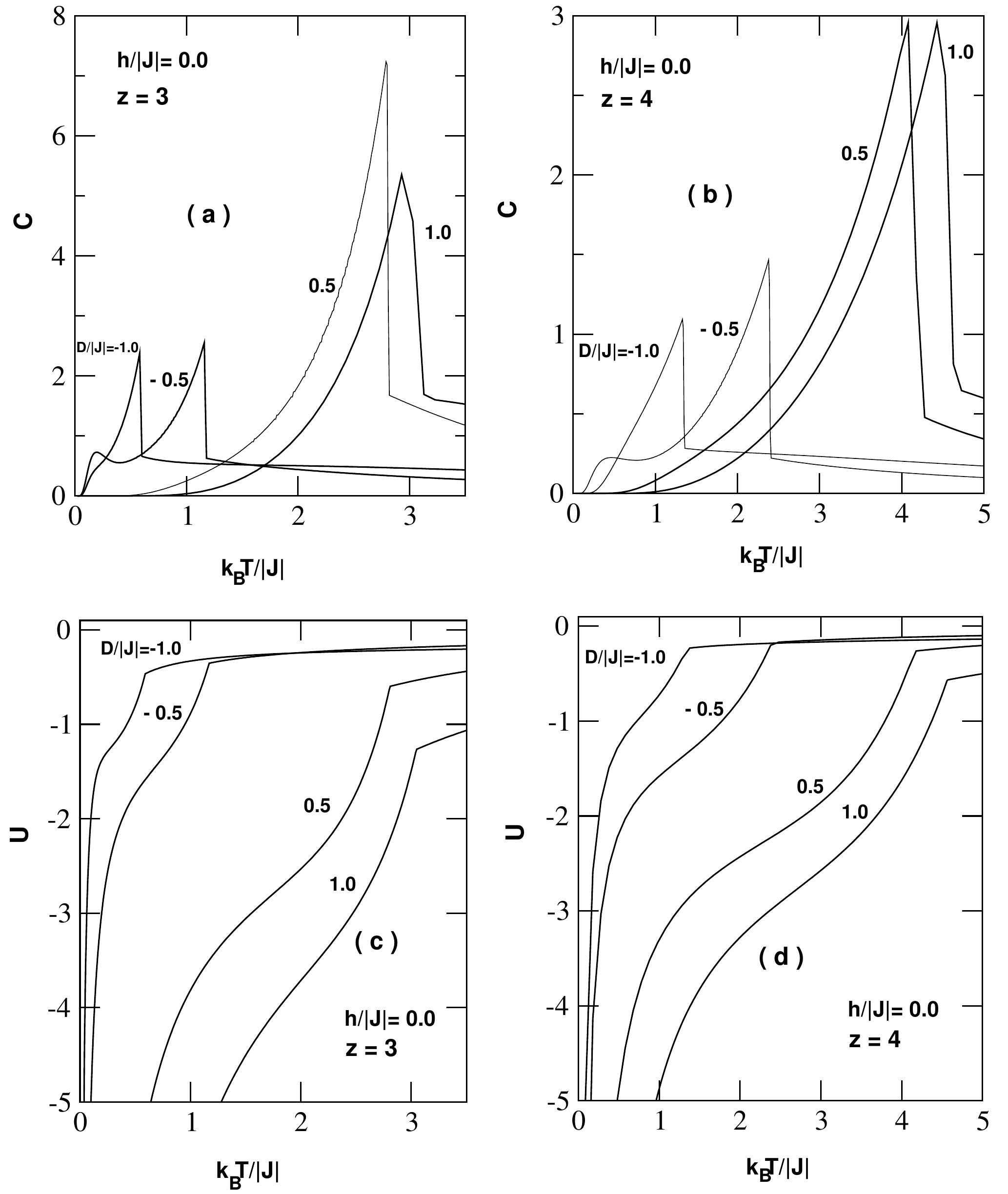}
	\end{center}
     \vspace{-4mm}
	\caption{Thermal variations of the specific heat and internal energy
		are calculated for $z=3, 4$ and selected values of the crystal-field $ D/|J| $ as shown in the 
		figures from panel (a) to panel (d). Values of the physical parameters
		considered are indicated in different panels. Analysis of the data in
		different panels  shows that the model only exhibits second-order transition for $z=3, 4$.}
	\label{figure5}
\end{figure} 

  Figure~\ref{figure4} illustrates some thermal variations of the sublattice magnetizations
  $M_{1/2}$ and $M_{7/2}$ when $z=3, 4$ and $6$ for selected values of the crystal-field $D/|J|$.
   The results are in perfect agreement with the
  ground-state phase diagram concerning the saturation values. Indeed, $M_{1/2}$ falls from 
  its unique saturation value~$\mp\frac{1}{2}$ with an increasing temperature whereas $M_{7/2}$ shows
  seven saturation values. The behaviours of the sublattice
  magnetizations $M_{1/2}$ and $M_{7/2}$ are quite similar. Moreover, one can notice that all the curves are 
  continuous and the Curie temperature $T_\text c$ at which both magnetization curves go to zero increases with 
  the crystal-field $D/|J|$ and the coordination number $z$.

  In figure~\ref{figure5}, we have plotted the thermal variations of the specific heat and the 
  internal energy for various values of the crystal-field as indicated in the figure. Both the specific heat and 
  the internal energy rapidly increase with increasing temperature and  make peak without jump discontinuities at 
  the same~$T_{\text{c}}$. By increasing the strength of the crystal-field and the coordination number,
  the value of $ T_{\text{c}}$ at which the transition occurs, increases and this can be easily 
  observed by comparing the results from different panels of figure~\ref{figure5}. The results obtained in this figure  confirm that
  the model only presents second-order transition for all values of the coordination number $z$.
   \begin{figure}[!t]
	\begin{center}
		\includegraphics[angle=0,width=0.7\textwidth]{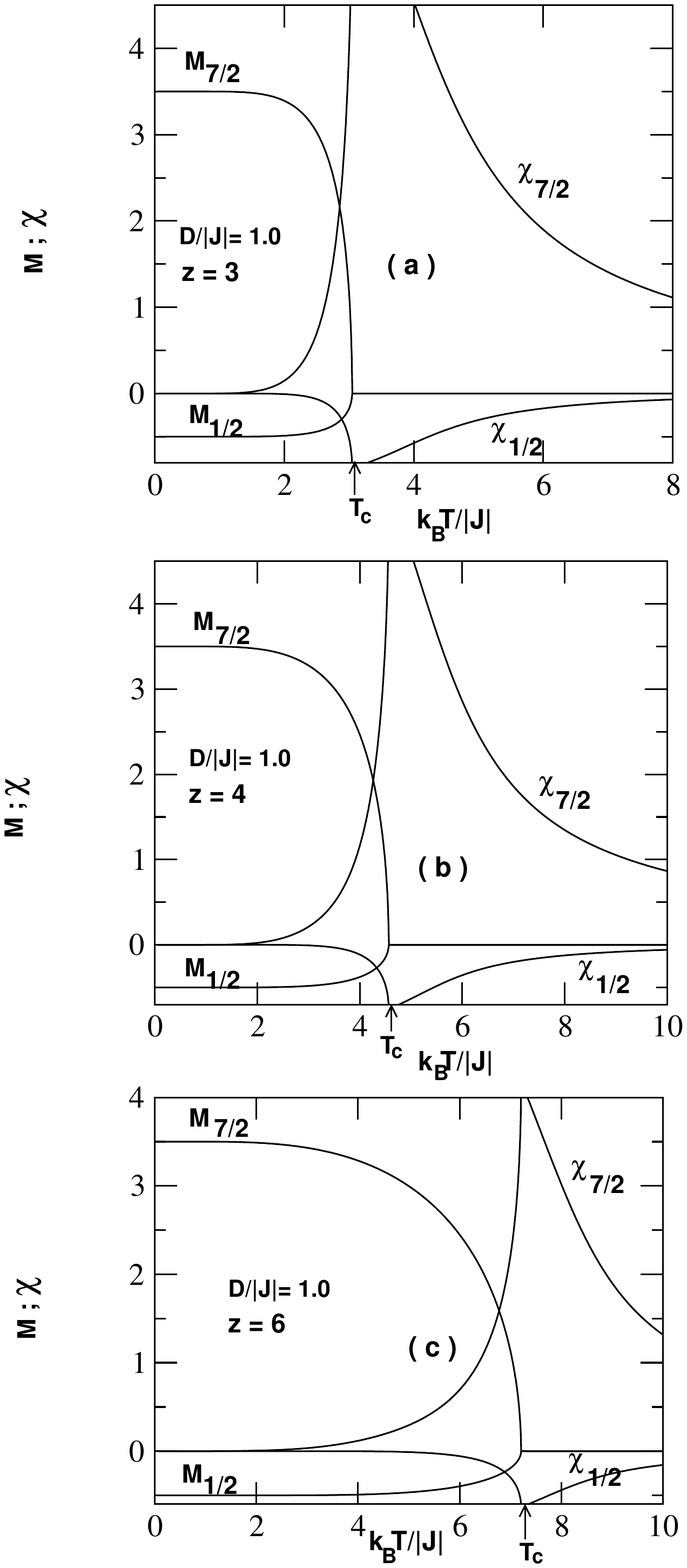}
	\end{center}
    \vspace{-3mm}
	\caption{The behaviour of the sublattice magnetizations and magnetic susceptibilities as a function of 
		temperature on the Bethe lattice  for $z = 3, 4$ and $6$ when
		$D/|J| = 1.0$.  Values of the physical parameters
		considered for the system are indicated in different panels.}
	\label{figure6}
\end{figure}  

  In figure~\ref{figure6}, we also present the temperature dependences of both  sublattice 
  magnetizations and susceptibilities when $z=3, 4, 6$ and $D/|J|=1$. From this figure,
   the value of the transition temperature $T_{\text{c}}$ increases with an increasing coordination number $z$. Here, $T_{\text{c}}$ separates the
  ferrimagnetic phase $\mathbf{F} \big(\pm{\frac{1}{2}}, \mp{\frac{7}{2}}\big)$ from the paramagnetic phase $(P)$ and $T_{\text{c}}/|J|=3.110$ 
  (respectively $T_{\text{c}}/|J|=4.644$ and $7.313$)
  for $z=3$ (respectively  for $z=4$ and $6$). Furthermore, one remarks that for 
  $T\rightarrow T_{\text{c}}$, $\chi_{7/2} \rightarrow +\infty$  whereas  $\chi_{1/2} \rightarrow -\infty$.
   For $T> T_{\text{c}}$, the susceptibility $\chi_{1/2}$ rapidly increases whereas the susceptibility $\chi_{7/2}$
   rapidly decreases when the temperature increases and is very far from the Curie temperature $T_{\text{c}}$,
   $\chi_{7/2} \rightarrow 0$ and  $\chi_{1/2} \rightarrow 0$.
\begin{figure}[!t]
	\begin{center}
		\includegraphics[angle=0,width=0.7\textwidth]{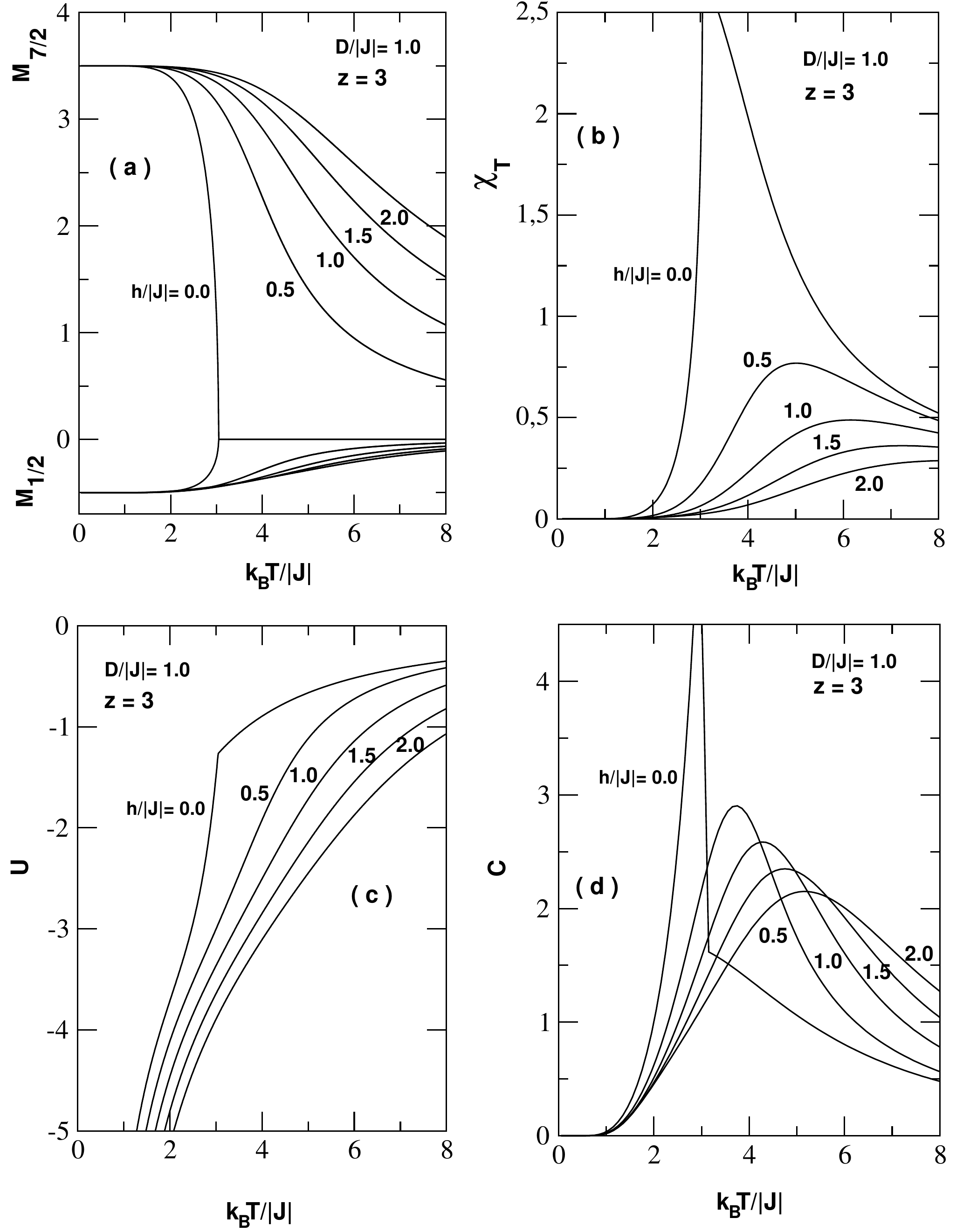}
	\end{center}
    \vspace{-4mm}
	\caption{The temperature dependence of the sublattice magnetizations $M_{1/2}$ , $M_{7/2}$ [panel (a)], the 
		total susceptibility 
		$\chi_\text{T}$ [panel (b)],
		the total internal energy $U$ [panel (c)] and the total specific heat $C$ [panel~(d)]. The following values of 
		the parameters are used: $ D/|J|= 1.0$;  
		$h / | J |$= $ 0.0; 0.5; 1.0; 1.5; 2.0$.}
	\label{figure7}
\end{figure}
   
  Let us now discuss the thermal variations of the sublattice magnetizations, the corresponding response functions
  and the internal energy of the system in the presence of the longitudinal magnetic field $h$.
  
  Figure~\ref{figure7} expresses the effects of an applied magnetic field $h$ on the magnetic properties of the model when
  $z=3$ and $D/|J|=1.0$ for selected values of $h/|J|$.
  In panel (a), the sublattice magnetizations continuously decrease from their saturation values 
  to non-zero values when the temperature increases. The remaining values of the sublattice magnetizations are 
  more important when the value of the applied magnetic field is high. Thus, one  observes that the system does not
  present any transition when $h/|J| \ne 0$. It is important to indicate that in the case of $h/|J|=0$, the model exhibits
  the second-order transition at a Curie temperature $T_\text{c}/|J|=3.110$, where the two sublattice magnetizations continuously
  go to zero after decreasing from their saturation values at $T=0$. In panels (b), (c) and (d), 
  we have displayed the temperature
  dependence of the total susceptibility $\chi_\text{T}$, the internal energy $U$ and the specific heat $C$,
   respectively. One can see from
  these panels that the response functions and the internal energy indicate a second-order
  transition which occurs at the same
 $T_\text{c}/|J|$ as in the case of $h/|J|=0$. For $h/|J| \ne 0$ and $T>T_{\text{c}}$, the response functions exhibit a maximum
 and the height of the maximum decreases when the value of the applied magnetic field increases.
  
\begin{figure}[!t]
\begin{center}
\includegraphics[angle=0,width=0.7\textwidth]{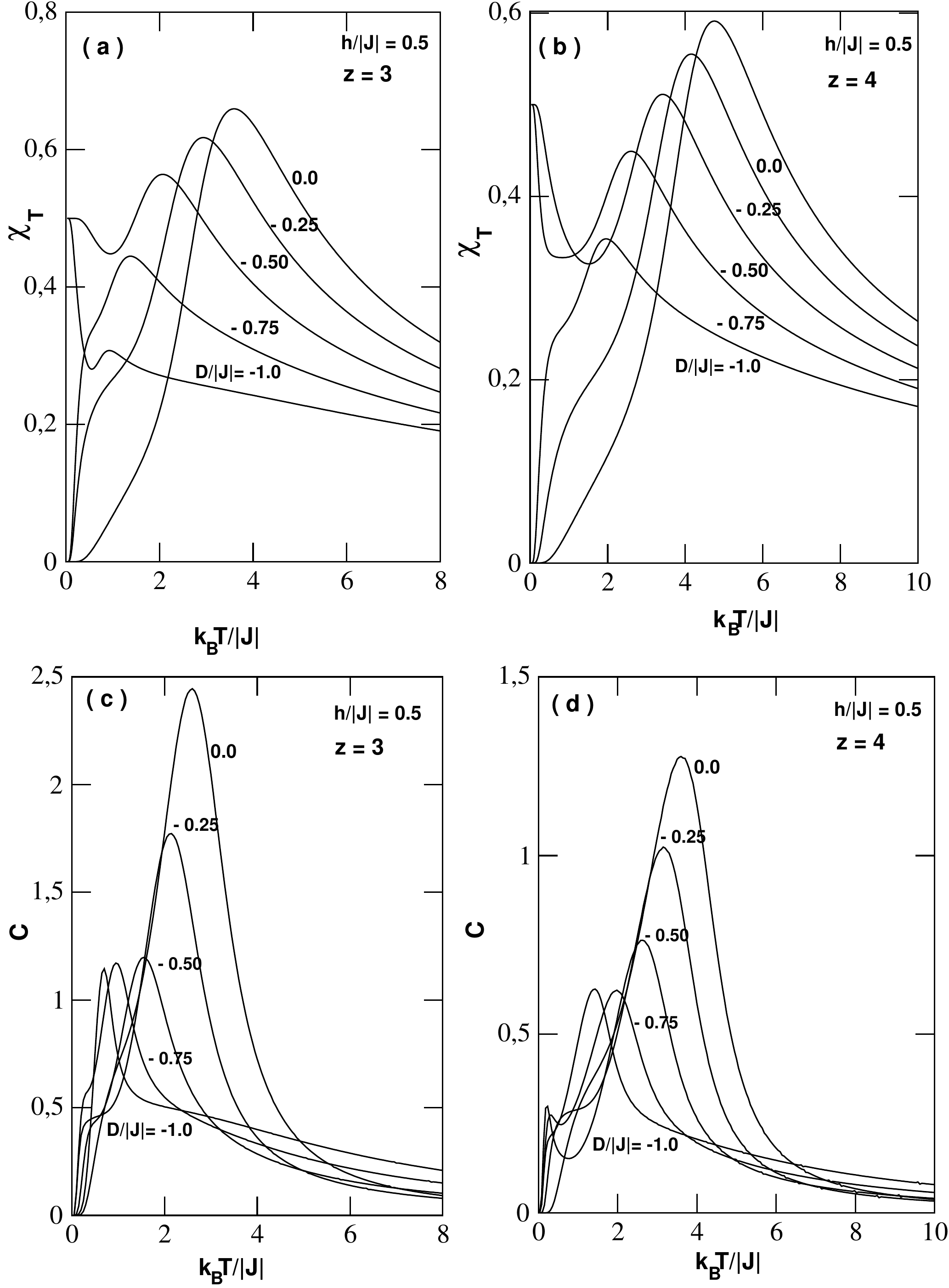}
\end{center}
\caption{Temperature variations of the response functions of the model at selected  
values of $D/|J|$ indicated on different curves illustrated for $z =3, 4$ and $h/|J|=0.5$.}
\label{figure8}
\end{figure}

 In figure~\ref{figure8}, we have presented the thermal variations of the
 response functions for some values of the system parameters to show the influence of $D/|J|$ on the system
 properties for $h/|J| \ne 0$. In the figure, one observes that the response functions show interesting
 behaviours. Indeed, the two studied response functions globally show a maximum at a certain value of the temperature.
 This temperature increases with the coordination number and the strength of the crystal-field. It is important 
 to mention that the height of the maximum of the two response functions also increases with increasing values of
 the strength of the crystal-field $D/|J|$ but the opposite holds when the coordination number $z$
 increases.
 
\begin{figure}[!t]
\begin{center}
\includegraphics[angle=0,width=0.74\textwidth]{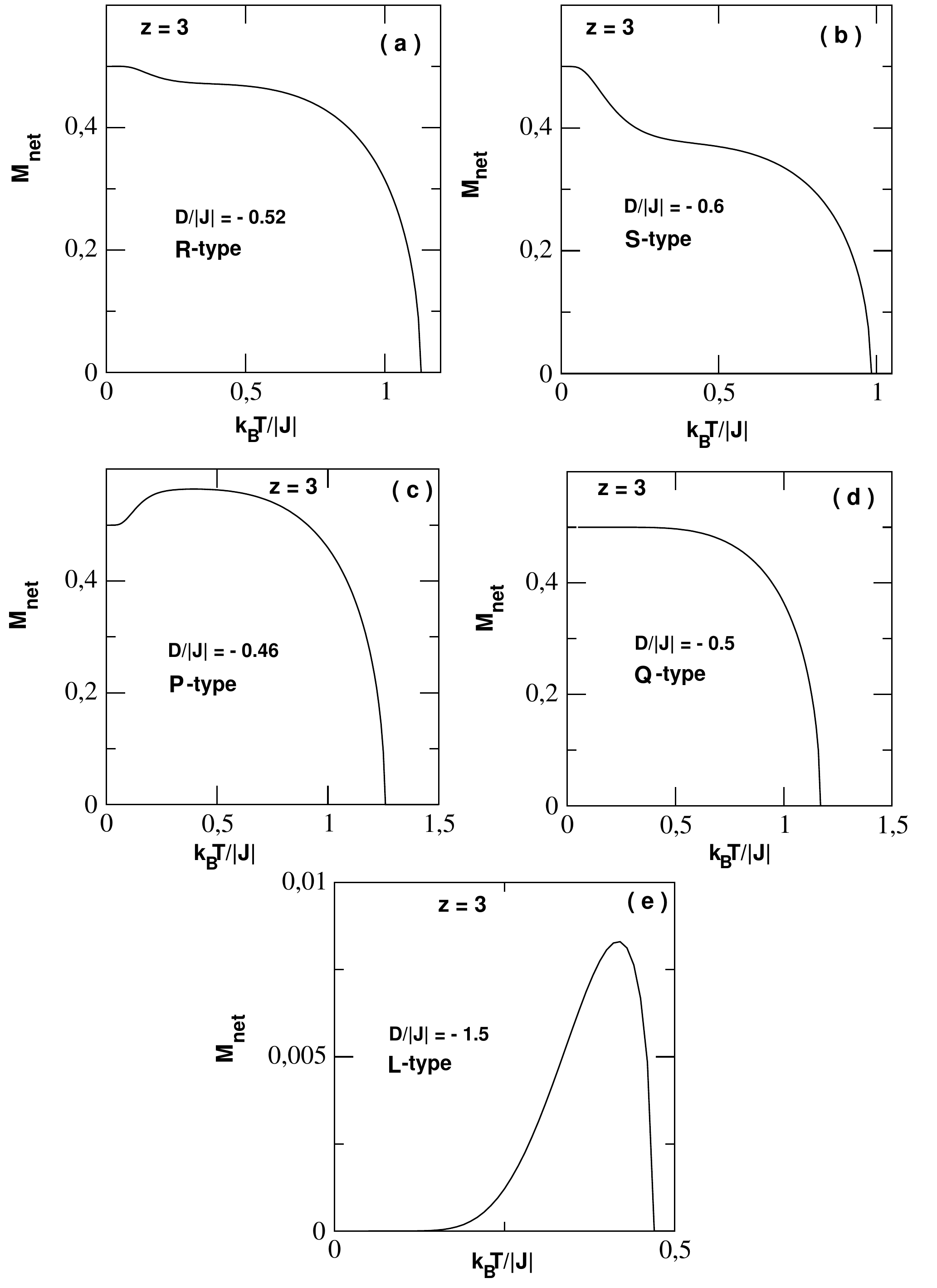}
\end{center}
\caption{The N\'{e}el nomenclature of average magnetization: (a) R-type for $D/|J|=-0.52$;  (b) S-type for $D/|J|=-0.6$;
 (c) P-type for $D/|J|=-0.46$;  (d) Q-type for $D/|J|=-0.5$ ; (e) R-type for $D/|J|=-1.5$.}
\label{figure9}
\end{figure}

  In figure~\ref{figure9}, we have investigated the global magnetization as a function of the temperature and obtained some 
compensation types of the model. The figure shows temperature dependencies of the global magnetization
$M_\text{net}$ for selected values of the crystal-field when $z=3$. As seen from figure~\ref{figure9}, the model 
exhibits five types of compensation behaviours, namely R-, S-, P-, Q- and L-type compensation behaviours as
classified in the extended N\'{e}el nomenclature \cite{la46a,la46b,la47,la48}.
 
\begin{figure}[!t]
\begin{center}
\includegraphics[angle=0,width=0.49\textwidth]{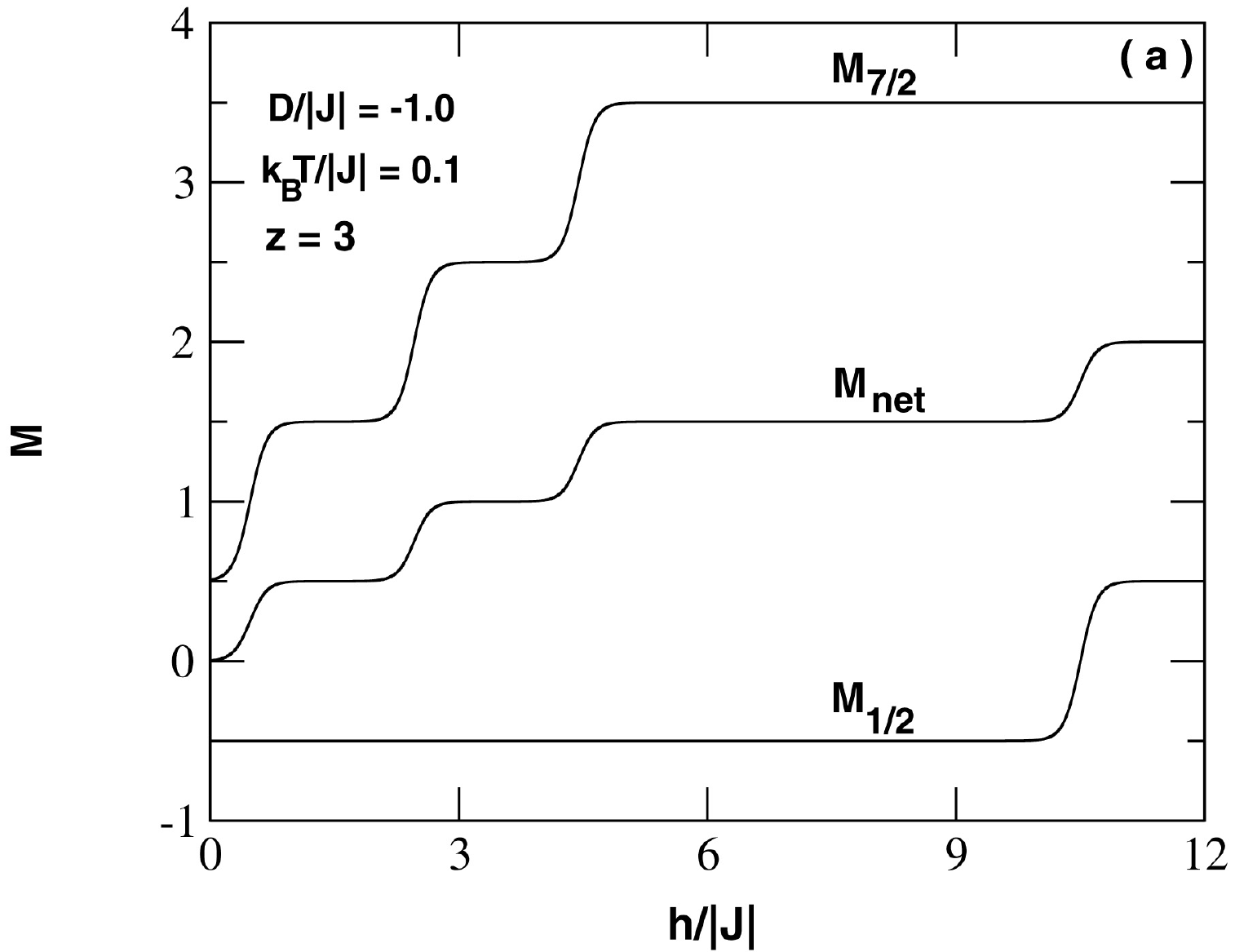}
\includegraphics[angle=0,width=0.49\textwidth]{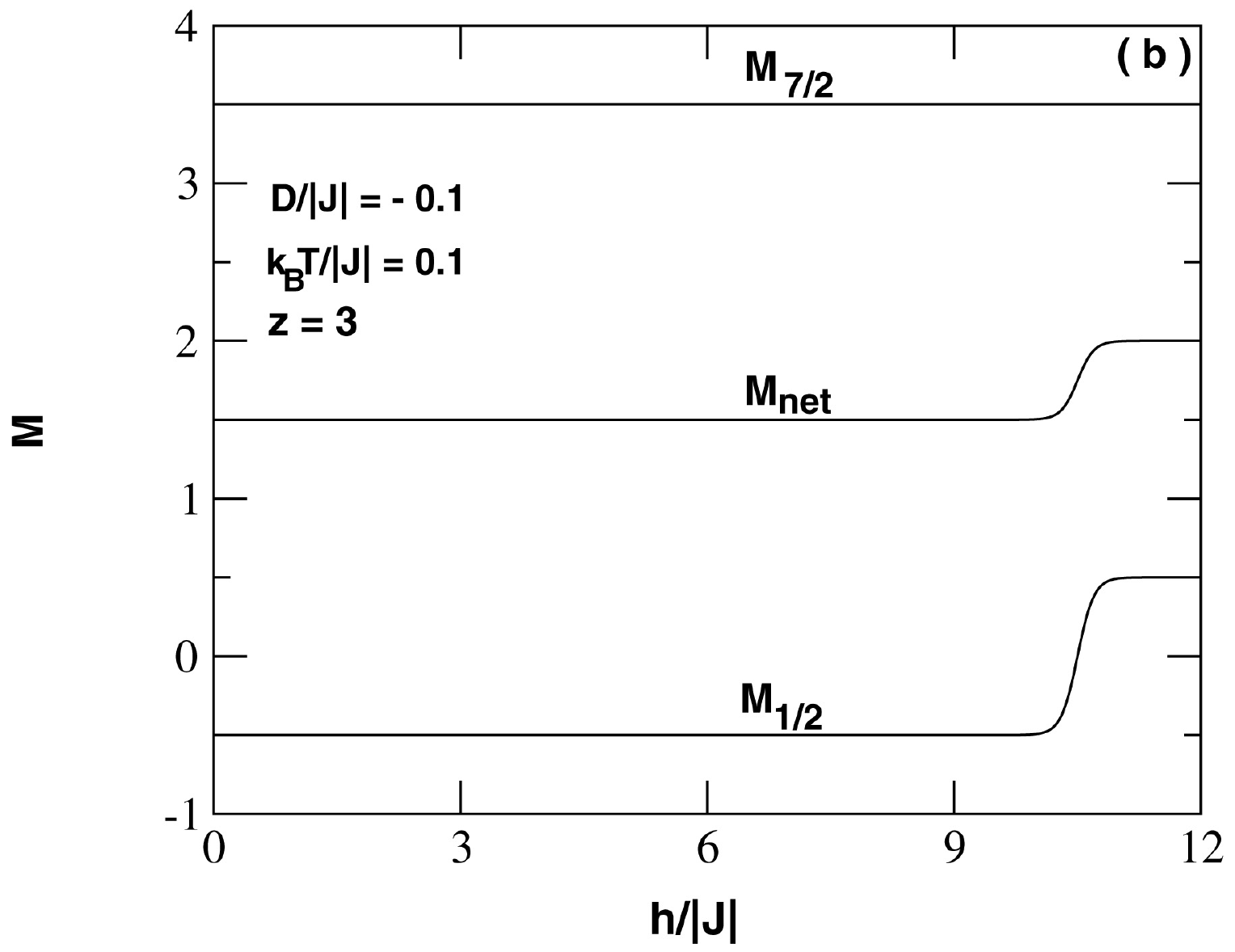}
\end{center}
\caption{ Magnetizations $M_{1/2}$, $M_{7/2}$ and $M_\text{net}$ plotted as functions of the magnetic field $h$ for selected values of the crystal-field
when $z=3$ as indicated in different panels.}
\label{figure10}
\end{figure}

  Moreover, we  investigate the low-temperature magnetic properties of the model. We plotted 
the sublattice magnetizations and the global magnetization at $k_\text{B} T/|J|=0.1$ for selected values
of the crystal-field as functions of the field $h$ as shown in figure~\ref{figure10}. In panel (a) where 
$D/|J|=-1$ and $z=3$, $M_\text{net}$ and $M_{7/2}$ respectively show five  and four step-like magnetization plateaus
 $(M_\text{net}=0, \frac{1}{2}, 1, \frac{3}{2}, 2)$ and $(M_{7/2}=\frac{1}{2}, \frac{3}{2}, \frac{5}{2}, \frac{7}{2})$ whereas
$M_{1/2}$ shows two step-like magnetization plateaus $(M_{1/2}=-\frac{1}{2}, \frac{1}{2})$.
 Also, from panel (b) where $D/|J|=0$ and $z=3$, only $M_\text{net}$ and $M_{1/2}$ present two step-like magnetization
plateaus \big($M_\text{net}=\frac{3}{2}, 2 $\big) and \big($M_{1/2}=-\frac{1}{2},\frac{1}{2}$\big). These obtained results 
are consistent with the ground-state phase diagram displayed in figure 2 of references \cite{la49,la50}.

  \section{Conclusion}
  In this paper, we have  studied the magnetic properties of the mixed spin-$\frac{1}{2}$ and
  spin-$\frac{7}{2}$ Ising ferrimagnetic model on the BL 
  in the presence of a longitudinal magnetic field by means of the recursion relations
  method. All the thermodynamical quantities of interest are calculated as functions of recursion relations.
  
  The ground-state phase diagram of the model is displayed as shown in figure~\ref{figure2}. From this 
  phase diagram, we have found eight existing and stable phases and along the ${D}/{q|J|}$-axis, three particular 
  hybrid phases appear at the three multicritical points $A_5$, $A_6$ and $A_7$.  The ground-state phase diagram
  is considered and used as a guide for obtaining different temperature phase diagrams. We also 
  investigated the phase diagrams in the $( D/|J|, k_{\text{B}}T/|J|)$ plane, shown in figure~\ref{figure3}. Then, in the presence and without the longitudinal magnetic field $h$, we examined the thermal variations of the
  sublattice magnetizations, the global magnetization, the corresponding response
  functions and the internal energy as reported in figures~\ref{figure4}--\ref{figure10}. From these 
  figures, the order-parameters in most cases showed a usual decay with thermal fluctuations. 
  By using these behaviours  and  the analysis of the corresponding
  response functions and the internal energy, the nature of  different
  phase transitions encountered is identified.  The model shows rich physical properties, namely
  the second-order transition and multicritical points  for all values 
  of the crystal-field interactions and for all values of the coordination number $z$.
  
  As a final note, it is useful to mention that different results achieved here
  are compared to those reported
  is some previous works \cite{la24,la45} and topological similarities are found.

\ukrainianpart

\title[]{ Дослідження методом ґратки Бете змішаної спін-$\frac{1}{2}$ та спін-$\frac{7}{2}$ моделі Iзінга в поздовжньому магнітному полі}

\author[S. Eddahri \textit{et al.}]{ С.~Еддагрі\refaddr{label4}, M.~Каріму\refaddr{label1}, A.~Разук\refaddr{label4,label5}, Ф.~Гонтінітде\refaddr{label1,label2},
	A.~Бенюссеф\refaddr{label3,label6} }
\addresses{
	\addr{label1} Інститут математики і фізичних наук (IMSP), Республіка Бенін\\
	\addr{label2} Університет Абомей-Калаві, фізичний факультет, Республіка Бенін\\
	\addr{label3} LMPHE, Факультет природничих наук, університет Мохаммеда  V, Рабат, Марокко\\
	\addr{label4} Лабораторія фізики матеріалів, факультет  наук і технологій, університет султана Муле Слімана, Марокко\\
	\addr{label5} Відділення фізики, полідисциплінарний факультет, університет султана Муле Слімана, Марокко\\
	\addr{label6} Академія наук і технологій Хассана II, Рабат, Марокко}

\makeukrtitle

\begin{abstract}
	Досліджено магнітні властивості змішаної спін-$\frac{1}{2}$ та спін-$\frac{7}{2}$ моделі Ізінга з кристалічним полем у поздовжньому магнiтному полі на ґратці Браве з використанням  точних рекурсивних співвідношень. Побудована фазова діаграма основного стану. Температурно-залежна фазова діаграма продемонстрована для випадку однорідного кристалічного поля на    площині $(k_\text{B}T/|J|, D/|J|)$ при відсутності зовнішнього обмеження для координаційних чисел $z = 3, 4$, $6$. Параметр порядку, відповідна функція відгуку і  внутрішня енергія обчислені та вивчені детально для встановлення справжньої природи фазових меж і відповідних температур. Термічні зміни 
	середньої намагніченості прокласифіковано відповідно до номенклатури Нееля.
	
	\keywords модель Ізінга,  намагніченість, функція відгуку, вільна енергія, фазова діаграма, перехід другого роду
\end{abstract} 


\newpage

\begin{thebibliography}{99}
	\bibitem{la1} White R.M., Science, 1985, \textbf{229}, 11,
	\bibdoi{10.1126/science.229.4708.11}.
	\bibitem{la2} Wood R., Understanding Magnetism, Tab Books Inc.: Blue Ridge Summit, Pennsylvania, 1988.
	\bibitem{la3}  K\"{o}ster E., J. Magn. Magn. Mater., 1988, \textbf{120}, 1,
	\bibdoi{10.1016/0304-8853(93)91274-B}.
	\bibitem{la4} Lueck L.B., Gilson R.G., J. Magn. Magn. Matter., 1990, \textbf{88}, 227, \bibdoi{10.1016/S0304-8853(97)90032-9}.
	\bibitem{la5} Itoh K., Kinoshita M., Molecular Magnetism: New Magnetic Materials, Kodansha and Gordon \& Breach Science Publishers, Tokyo--Amsterdam, 2000. 
	\bibitem{la6} Linert W., Verdaguer M. (Eds.), Molecular Magnets: Recent Highlights, Springer, Berlin, 2003.
	\bibitem{la7} Gatteschi D., Adv. Mater., 1994, \textbf{6}, 635, 
	\bibdoi{10.1002/adma.19940060903}.
	\bibitem{la8} Miller J.S., Epstein A.J., Chem. Eng. News, 1995, \textbf{73}, 30,
	\bibdoi{10.1021/cen-v073n040.p030}.
	\bibitem{la9} Kahn O., Molecular Magnetism, VCH Publishers, Inc., New York, 1993.
	\bibitem{la10} Thompson C.J., Mathematical Statistical Mechanics, Princeton University Press, Princeton, New Jersey, 1992.
	\bibitem{la11} Bob\'ak A., Physica A, 1998, \textbf{258}, 140, \bibdoi{10.1016/S0378-4371(98)00233-7}.
	\bibitem{la12} Quadros S.G.A., Salinas S.R., Physica A, 1994, \textbf{206}, 479, \bibdoi{10.1016/0378-4371(94)90319-0}. 
	\bibitem{la13}   El Bouziani M., Gaye A.,  Jellal A., Physica A, 2013, \textbf{392}, 689, \bibdoi{10.1016/j.physa.2012.10.007}.
	\bibitem{la14} Da Cruz Filho J.S.,  Godoy M.,  Arruda A.S., Physica A, 2013, \textbf{392}, 6247, \bibdoi{10.1016/j.physa.2013.08.007}.
	\bibitem{la15} Miao H.,  Wei G.,  Geng J., J. Magn. Magn. Mater., 2009,
	\textbf{321}, 4139, \bibdoi{10.1016/j.jmmm.2009.08.018}.
	\bibitem{la16} Mohamad H.K., Domashevskaya E.P., Klinskikh A.F., Physica A, 2009, \textbf{388}, 4713, \\ \bibdoi{10.1016/j.physa.2009.08.014}.
	\bibitem{la17} Mohamad H.K., Int. J. Adv. Res., 2014, \textbf{2}, 442. 
	\bibitem{la18} Kaneyoshi T., Chen J.C., J. Magn. Magn. Mater., \textbf{98}, 1991, 201, \bibdoi{10.1016/0304-8853(91)90444-F}.
	\bibitem{la19}Benyoussef A., El Kenz A., Kaneyoshi T., J. Magn. Magn. Mater., 1994, \textbf{131}, 179, \\ \bibdoi{10.1016/0304-8853(94)90026-4}.
	\bibitem{la20} Bob\'{a}k A., Jur\v{c}i\v{s}in M., Physica A, 1997, \textbf{240}, 647, \bibdoi{10.1016/S0378-4371(97)00044-7}.
	\bibitem{la21} De Oliveira D.C., Silva A.A.P., de Albuquerque D.F., de Arruda~A.S., Physica A, 2007, \textbf{386}, 205, \\
	\bibdoi{10.1016/j.physa.2007.07.073}.
	\bibitem{la22} Kaneyoshi T., Physica A, 1994, \textbf{205}, 677, \bibdoi{10.1016/0378-4371(94)90229-1}.
	\bibitem{la23} Guo K.-T., Xiang S.-H., Xu H.-Y., Li X.-L., Quantum Inf. Process., 2014, \textbf{13}, 1511, \\ \bibdoi{10.1007/s11128-014-0745-7}.
	\bibitem{la24} Deviren B., Keskin M., Canko O., Physica A, 2009, \textbf{388}, 1835, \bibdoi{10.1016/j.physa.2009.01.032}.
	\bibitem{la25} Buend\'ia G.M.,  Liendo J.A., J. Phys.: Condens. Matter, 1997, \textbf{9}, 5439, \bibdoi{10.1088/0953-8984/9/25/011}.
	\bibitem{la26} Godoy M., Figueiredo W., Phys. Rev. E, 2002, \textbf{66}, 036131,  \bibdoi{10.1103/PhysRevE.66.036131}.
	\bibitem{la27} Cambu\'i D.S., de Arruda A.S.,  Godoy M.,
	Int. J. Mod. Phys. C, 2012, \textbf{23}, 1240015, \\ \bibdoi{10.1142/S0129183112400153}.
	\bibitem{la28} Feraoun A., Zaim A., Kerouad M., Physica B, 2014, \textbf{445}, 74, \bibdoi{10.1016/j.physb.2014.03.071}.
	\bibitem{la29} \v{Z}ukovi\v{c} M., Bob\'{a}k A., J. Magn. Magn. Mater., 2010, \textbf{322}, 2868, \bibdoi{10.1016/j.jmmm.2010.04.043}.
	\bibitem{la30} Jiang W., Bai B.-D., Phys. Status Solidi B, 2006, \textbf{243}, 2892, \bibdoi{10.1002/pssb.200541244}.
	\bibitem{la31} Essaoudi I., B\"{a}rner  K., Ainane A., Saber M., Physica A, 2007, \textbf{385}, 208, \bibdoi{10.1016/j.physa.2007.06.037}.
	\bibitem{la32} Yessoufou R.A.,  Bekhechi S.,  Hontinfinde F., Eur. Phys. J. B, 2011, \textbf{81}, 137, \bibdoi{10.1140/epjb/e2011-10825-7}.
	\bibitem{la33} Kpl\'e J.,  Yessoufou R.A.,  Hontinfinde F.,  Afr. Rev. Phys., 2012, \textbf{7}, 319.
	\bibitem{la34} Yigit A., Albayrak E., Chin. Phys. B, 2012, \textbf{21}, 020511, \bibdoi{10.1088/1674-1056/21/2/020511}.
	\bibitem{la35} Ekiz C., Phys. Lett. A, 2007, \textbf{367}, 483, \bibdoi{10.1016/j.physleta.2007.03.038}.
	\bibitem{la36} Albayrak E.,  Yigit A.,  Phys. Lett. A, 2006, \textbf{353}, 121, \bibdoi{10.1016/j.physleta.2005.12.077}.
	\bibitem{la37}  Karimou M.,  Yessoufou R.,  Hontinfinde F., Int. J. Mod. Phys. B, 2015, \textbf{29}, 1550194, \\ \bibdoi{10.1142/S0217979215501945}.
	\bibitem{la38}  Ekiz C., J. Magn. Magn. Mater., 2006,  \textbf{307}, 139, \bibdoi{10.1016/j.jmmm.2006.03.059}.
	\bibitem{la39} Ekiz C., Commun. Theor. Phys., 2009, \textbf{52}, 539, \bibdoi{10.1088/0253-6102/52/3/30}.
	\bibitem{la40} Stre\v{c}ka J., Ja\v{s}\v{c}ur M., Acta Phys. Slovaca, 2015, \textbf{65}, 235, [Preprint \arxiv{1511.03031v2}, 2015].
	\bibitem{la41} Ja\v{s}\v{c}ur M., \v{S}tub\v{n}a V., Sza\l{}owski K., Balcerzak T., J. Magn. Magn. Mater., 2016, \textbf{417}, 92, \\
	\bibdoi{10.1016/j.jmmm.2016.05.048}.
	\bibitem{la42} Koyama K., Fujii H., Goto T., Fukuda H., Janssen Y., Physica B, 2001, \textbf{294--295}, 168, \\ \bibdoi{10.1016/S0921-4526(00)00634-7}.
	\bibitem{la43} Albertini F., Bolzoni F., Paoluzi A., Pareti L., Zannoni E., Physica B, 2001, \textbf{294}, 172, \\ 
	\bibdoi{10.1016/S0921-4526(00)00635-9}.
	\bibitem{la44} Baxter R.J., Exactly Solvable Models in Statistical Mechanics, Academic Press, London, 1982.
	\bibitem{la46a} N\'{e}el L., Ann. Phys. Paris, 1932, \textbf{18}, 5.
	\bibitem{la46b} N\'{e}el L., Ann. Phys. Paris, 1948, \textbf{3}, 137.
	\bibitem{la47}  Ekiz C., Stre\v{c}ka J., Ja\v{s}\v{c}ur M., Cent. Eur. J. Phys., 2009,  \textbf{7}, 509,
	\bibdoi{10.2478/s11534-009-0043-7}.
	\bibitem{la48} Chikazumi S., Physics of Ferromagnetism, Oxford University Press, Oxford, 1997.
	\bibitem{la49} Hovhannisyan V.V., Stre\v{c}ka J., Ananikian N.S., J. Phys.: Condens. Matter, 2016, \textbf{28}, 8, \\ \bibdoi{10.1088/0953-8984/28/8/085401}.
	\bibitem{la50} Yao X., Dong S., Yu H., Liu J., Phys. Rev. B, 2006, \textbf{74}, 134421, \bibdoi{10.1103/PhysRevB.74.134421}.
	\bibitem{la45} Karimou M., Yessoufou R.A., Oke T.D., Kpadonou A., Hontinfinde F., Condens. Matter
	Phys., 2016, \textbf{19}, 33003, \\ \bibdoi{10.5488/CMP.19.33003}. 
\end{thebibliography}
\end{document}